\documentclass[floatfix,twocolumn,nofootinbib,superscriptaddress,a4paper,preprintnumbers]{revtex4-1}

\usepackage{amsmath}
\usepackage{amssymb}
\usepackage{graphicx}
\usepackage[T1]{fontenc}
\usepackage{slashed}
\usepackage[utf8]{inputenc}
\usepackage{xspace}
\usepackage{color}
\usepackage{ulem}

\newcommand{\GeV}{{\ensuremath{{\mathrm{GeV}}}}}

\newcommand\SARAH{{\tt SARAH}\xspace}

\newcommand\SPheno{{\tt SPheno}\xspace}

\newcommand\Mathematica{{\tt Mathematica}\xspace}

\newcommand{\AddrPeking}{Center for High-Energy
Physics, Peking University, Beijing, 100871, P. R. China}
\newcommand{\AddrKavli}{State Key Laboratory of Theoretical Physics
and Kavli Institute for Theoretical Physics, China (KITPC),
Institute of Theoretical Physics, Chinese Academy of Sciences,
Beijing 100190, P. R. China}
\newcommand{\AddrChengdu}{School of Physical Electronics,
University of Electronic Science and Technology of China,
Chengdu 610054, P. R. China}
\newcommand{\AddrCERN}{Theory Division, CERN, 1211 Geneva 23, Switzerland}
\newcommand{\AddrBeijing}{Institute of Physics Chinese Academy of sciences, Beijing 100190, P. R. China}

\preprint{CERN-PH-TH-2015-237}
\begin{document}

\hfill{CERN-PH-TH-2015-237}

\title{Supersoft Supersymmetry, Conformal Sequestering,
and Single Scale Supersymmetry Breaking}

\author{Ran Ding}
%\email{dingran@mail.nankai.edu.cn}
\affiliation{\AddrPeking}

\author{Tianjun Li}
%\email{tli@itp.ac.cn}
\affiliation{\AddrKavli}
\affiliation{\AddrChengdu}

\author{Florian Staub}
%\email{florian.staub@cern.ch}
\affiliation{\AddrCERN}

\author{Bin Zhu}
%\email{zhubin@mail.nankai.edu.cn}
\affiliation{\AddrBeijing}

\begin{abstract}

Supersymmetric Standard Models (SSMs) with Dirac gauginos have the
appealing supersoft property that they only cause finite contributions to
scalar masses. Considering gauge mediated SUSY breaking with conformal
sequestering and assuming there is one and only one fundamental
parameter with dimension mass arising from supersymmetry breaking,
we find a cancellation between the dominant terms that contribute
to the EWFT. The resulting EWFT measure can be of order one even for
supersymmetric particle masses and $\mu$-terms in the TeV range.
\end{abstract}
\maketitle

% =============================================================================
%\section{Introduction}
%\label{sec:intro}
% =============================================================================

{\bf Introduction}--Supersymmetry (SUSY) provides a natural solution to the gauge
hierarchy problem in the Standard Model (SM). In the Supersymmetric SMs (SSMs)
with $R$ parity, gauge coupling unification can be achieved, and the Lightest
Supersymmetric Particle (LSP) is a dark matter candidate. However, after the first run
of the Large Hadron Collider (LHC), the former top candidate for physics
beyond the SM, the Minimal SSM (MSSM), has lost a lot of its attraction. One reason is
the discovery of the SM-like Higgs boson with a mass of  $125$~GeV~\cite{Chatrchyan:2012ufa,Aad:2012tfa}.
In order to obtain the correct Higgs mass, there are two possibilities in 
the MSSM: either there
must be a very large mixing among the supersymmetric partners of top quarks, or
the SUSY breaking soft masses must be much heavier than naively expected. The first possibility
is often disfavoured by charge and colour breaking minima~\cite{Camargo-Molina:2013sta, Blinov:2013fta,
  Chowdhury:2013dka, Camargo-Molina:2014pwa, Chattopadhyay:2014gfa}, while the second one raises
the question if the MSSM is really a natural solution to gauge hierarchy problem. This has
caused an increasing interest in non-minimal SUSY models. The focus was mainly on models
which enhance the Higgs at tree level to reduce the fine-tuning (FT)~\cite{BasteroGil:2000bw,
  Dermisek:2005gg,Zhang:2008jm,Ellwanger:2011mu,Ross:2011xv,%
  Hirsch:2011hg,Ross:2012nr,Gherghetta:2012gb,Perelstein:2012qg,Kim:2013uxa,Kaminska:2014wia,Binjonaid:2014oga}.
In addition, the other ideas like $R$-symmetric SSMs with Dirac instead of Majorana gauginos
became much more popular in the last few years~\cite{Fox:2002bu,Chacko:2004mi,Carpenter:2005tz,Antoniadis:2006eb,
Kribs:2007ac,Amigo:2008rc,Benakli:2008pg,Benakli:2009mk,Benakli:2010gi,
Benakli:2011vb,Choi:2010an,Abel:2011dc,Benakli:2011kz,Heikinheimo:2011fk,
Kribs:2012gx,Kalinowski:2011zzc,Davies:2012vu,Goodsell:2012fm,Benakli:2012cy,
Abel:2013kha,Kribs:2013oda,Csaki:2013fla,Busbridge:2014sha,Benakli:2014cia,
Chakraborty:2014sda,Ding:2015wma,Nelson:2015cea,Alves:2015bba,Chakraborty:2015wga,Goodsell:2015ura,Martin:2015eca}.
On the one hand, such models are known to be supersoft since they only give finite contributions
to scalar masses \cite{Fox:2002bu, Kribs:2012gx}. On the other hand, they can reduce existing mass limits
from SUSY searches and weaken bounds from flavour physics \cite{Kribs:2007ac}. It is somehow surprising,
but the electroweak FT (EWFT) question in the SSMs with Dirac gauginos and a specific SUSY breaking mechanism
has not been addressed so far. So we shall close this gap here.

The single scale SUSY provides an elegant solution to
the SUSY EWFT problem~\cite{Leggett:2014mza, Leggett:2014hha, Du:2015una}.
In particular, the original conditions for string inspired SSMs are mainly~\cite{Du:2015una}:
(1) The K\"ahler potential and superpotential can be calculated in principle or at least inspired
from a fundamental theory such as string theory with suitable compactifications;
(2) There is one and only one chiral superfield which breaks supersymmetry;
(3) All the mass parameters in the SSMs must arise from supersymmetry breaking. With these conditions,
one can show that the SUSY EWFT measure is automatically of order one. The above conditions seem to be too strong,
thus,  we point out that the essential condition is: there is one and only one
fundamental mass parameter
and the coefficients to set the different mass scales to be determined. Or simply speaking,
all dimensionful parameters in the SSMs are correlated. In particular, all dimensionful parameters
can be further relaxed to all the dimensional parameters with large EWFT measures,
and we shall call it the effective single scale condition.

In the minimal $R$-symmetric SSM (MRSSM) with Dirac gauginos, we present for the first time 
the SUSY breaking soft terms from  gauge mediated SUSY breaking (GMSB)~\cite{Benakli:2010gi}
with conformal sequestering~\cite{Luty:2001jh, Luty:2001zv, Murayama:2007ge}, and find
that the naive EWFT measure turns out to be similar to the other SSMs,
except the minor improvements due to supersoft property
and additional loop contributions to the Higgs boson mass. With our above updated condition
for single scale SUSY, we show a perfect cancellation analytically and numerically between the
  dominant terms that contribute to the EWFT. The resulting EWFT measure
  can be of order one even for the supersymmetric particle (sparticle) masses in the TeV range.
In particular, it is not
necessary that the dimensionful parameters in the superpotential have to be tuned to be small
as this is usually the case. In a wide range
of the parameter space we find a precise cancellation among different contributions to the
EWFT measures.

{\bf The SUSY Breaking Soft Terms}--The generic new soft terms in the MRSSM are
\begin{align}
\mathcal{L}=(m_D\lambda_i\psi_{A_i}+b_A A^2+h.c.)+m_A^2 \left|A\right|^2~,~
\label{eqn:dirac}
\end{align}
where $\lambda$ is a gaugino,  $\psi$ and $A$ are the fermionic and scalar components of
a chiral adjoint superfield, $m_D$ is the Dirac gaugino mass, and
$b_A$ and $m_A^2$ are the holomorphic and,respectively, non-holomorphic masses.

In the simplest ansatz that the origin of the Dirac mass term is the operator
for gauge field strengths $W_{\alpha}^{\prime}$ and $W_{j}^{\alpha}$
\begin{align}
W_{ssoft}=\frac{W_{\alpha}^{\prime}W_{j}^{\alpha}A_j}{\Lambda}~,~
\end{align}
a massless scalar in the adjoint representation is predicted~\cite{Csaki:2013fla}. This observation has triggered efforts in constructing
phenomenological reliable models with Dirac gauginos \cite{Carpenter:2015mna,Alves:2015bba,Alves:2015kia}. In general,
the aim is to get $m_D^2\sim m_A^2\sim b_A$.
However, if $b_A$ and $m_D$ are generated at one loop,
$b_A$ is naturally larger than $m_D^2$ by a loop factor of $16\pi^2$.
To address this $m_D-b_A$ problem and generate
the proper Dirac gaugino and scalar masses, we introduce two pairs of messenger fields
for the GMSB~\cite{Benakli:2010gi}
and consider the
conformal sequestering~\cite{Luty:2001jh, Luty:2001zv, Murayama:2007ge}.
Supposing the hidden sector interactions are strong below
the messenger scale $M_{\text{mess}}$ down to some scale where conformality is broken, we obtain
\begin{align}
 m_{D_i} =& \frac{g_i}{16\pi^2} \frac{C_{D_i}\lambda_i}{6{\sqrt 2}} \frac{\Lambda^{\prime 2}_F}{M_{\text{mess}}}~,~ 
 b_{A_i} =  -\frac{1}{16\pi^2}  \frac{C_{b_i}\lambda^2_i}{2^{\delta_i}} \Lambda^{\prime 2}_F ~,~ \nonumber \\
 m^2_{A_i} = & \left(\frac{1}{32\pi^2}  \frac{\lambda^2_i}{2^{\delta_i}}+
 \frac{1}{128 \pi^4}  \sum_i C_i(A_i) g_i^4  \right) C_{A_i} \Lambda^{\prime 2}_F ~,~  \nonumber \\
 m^2_{\phi} = & \frac{1}{128 \pi^4}  \sum_i C_i(\phi) g_i^4  C_{\phi} \Lambda^{\prime 2}_F ~,~
\end{align}
where $(\delta_1,\delta_2, \delta_3)=(0, 1, 1)$,
$g_i$ and $\lambda_i$ are gauge and Yukawa couplings,
$\phi$ represents scalars not appearing in the adjoint representation,
$C_{D_i/b_i/A_i/\phi}$ is the conformal sequestering suppression factor, and
$C_i(A_i/\phi)$ is the quadratic Casimir index. For simplicity, we assume
$C_{b_i}=C_{A_i} =C_{\phi} \equiv C_{XX}$, and define
\begin{eqnarray}
  y_i ~\equiv~ \frac{C_{D_i}\lambda_i}{6{\sqrt 2}} \frac{\Lambda^{\prime 2}_F}{M_{\text{mess}} \Lambda_D}~,~
  \Lambda^2_F ~\equiv~ C_{XX} \Lambda^{\prime 2}_F~,~
\end{eqnarray}
where $\Lambda_D$ and $\Lambda_F$ are roughly the same mass scales.
Assuming that $C_{XX} << C_{D_i}$ and $10 \lambda_i \leq g^2_1/2{\sqrt 2}\pi$, we approximately have
\begin{align}
\label{eq:Boundary1}
m_{D_i} =& \frac{g_i y_i}{16\pi^2} \Lambda_D ~,~~~ b_{A_i} \simeq 0 ~,~\\
 m^2_{A_i/\phi} \simeq & \frac{1}{128 \pi^4}  \sum_i C_i(A_i/\phi) g_i^4  \Lambda^{ 2}_F ~.~
\label{eq:Boundary2}
\end{align}

{\bf The MRSSM}--The particle content of the MRSSM is the MSSM extended by adjoint superfields for
all gauge groups necessary to construct Dirac gaugino masses as well as by
two chiral iso-doublets $R_u$ and $R_d$ with $R$ charge $2$ to build $\mu$ like terms.
Thus, the superpotential is
\begin{align}
  \nonumber W = &  - Y_d \,\hat{d}\,\hat{q}\hat{H}_d\,- Y_e \,\hat{e}\,\hat{l}\hat{H}_d\,
  +Y_u\,\hat{u}\,\hat{q}\hat{H}_u + \mu_D\,\hat{R}_d \hat{H}_d\,
  \, \\ \nonumber &
  +\mu_U\,\hat{R}_u\hat{H}_u\,+\hat{S}(\lambda_d\,\hat{R}_d\hat{H}_d\,+\lambda_u\,\,\hat{R}_u\hat{H}_u) \\
 &+  \lambda^T_d\,\hat{R}_d \hat{T}\,\hat{H}_d\,+\lambda^T_u\,\hat{R}_u\hat{T}\,\hat{H}_u \,~.~\,
\label{eq:superpot}
 \end{align}
All the other terms are forbidden by the $R$-symmetry as the Majorana gaugino masses and trilinear soft-breaking couplings are. However, a soft-breaking term
$B_\mu$ necessary to give mass to the pseudoscalar Higgs is allowed by this symmetry.
The tree-level Higgs mass is even smaller than the MSSM because of negative
contributions from the new $D$-terms proportional to the Dirac gaugino masses. Moreover, the stops can not be used to push this mass significantly since all $A$-terms are forbidden by
 $R$-symmetry. Nevertheless, it has been shown that the large loop corrections stemming from the new superpotential terms $\lambda_i$ and $\lambda^T_i$ ($i=u,d$)
increase the Higgs mass to the demanded level \cite{Diessner:2014ksa,Diessner:2015yna}. Moreover, this model is consistent with gauge coupling unification \cite{Goodsell:2015ura}. Thus, it
is natural to embed it in a constrained SUSY breaking scenario.

We use the boundary conditions defined in Eqs.~(\ref{eq:Boundary1})--(\ref{eq:Boundary2})
in the limit $\Lambda_D/M  \to 0$ to calculate most
soft masses at the conformal scale $M$. Only the soft-mass for the singlet $m_s^2$ and $B_\mu$,
which can also be generated via Yukawa mediations, are derived from the
minimization conditions at the vacuum.
The other two minimization conditions are used to calculate $\mu_D$ and $\mu_U$. In short, we have
the following input parameters
\begin{eqnarray}
\label{eq:input}
& \Lambda_F, \, \Lambda_D, \, M,\, y_i ,\, \lambda_u,\, \lambda_d,\, \lambda^T_u, \, \lambda^T_d, \,
\tan\beta, \, v_s,\, v_T ~,~
\end{eqnarray}
where  $\tan\beta \equiv \langle H_u^0 \rangle/\langle H_d^0 \rangle$, and $v_s$ and $v_T$ are the Vacuum
Expectation Values (VEVs) of the singlet and neutral triplet.
Also, we assume $\mu_D$ and $\mu_U$ are positive.

%\section{Naturalness}
%\label{sec:FT}

{\bf Naturalness}--To quantize the EWFT size,  we adopt the measure introduced in Refs.~\cite{Ellis:1986yg, Barbieri:1987fn}
\begin{equation}
\label{eq:measure}
\Delta_{FT} \equiv  {\text{Max}}\{\Delta _{\alpha}\},\qquad \Delta _{\alpha}\equiv \left|\frac{\partial \ln
  M_Z^{2}}{\partial \ln \alpha} \right|  \;,
\end{equation}
where $\alpha$ is a set of independent parameters, and $\Delta_\alpha^{-1}$ gives an estimate of the accuracy to which
the parameter $\alpha$ must be tuned to get the correct electroweak symmetry breaking (EWSB)
scale \cite{Ghilencea:2012qk}. The smaller $\Delta_{FT}$, the more natural the model under consideration is. We use the conformal scale $M$  as a reference scale and calculate the FT with respect to $\{\Lambda_F, \Lambda_D, y_i, \lambda_{d,u}, \lambda^T_{d,u}, \mu_{D,U}, m_s^2, B_\mu\}$.

For large regions in parameter space, the main EWFT sources are   $\mu_U$ and
the scale $\Lambda_F$ because of the impact on the running soft mass $m_{H_u}^2$ responsible for EWSB.
If we only include the terms proportional to top Yukawa coupling in the running, we can estimate the EWFT measures
for these two parameters to be
\begin{align}
\label{eq:FT1}
 \Delta_{FT}(\Lambda_F) \approx& \left|\frac{ \sqrt{2} \Lambda^2_F}{384 \pi^4 v^2}  \left((32 (-1 + R) + 9 (1 + R) g_2^4)) \right)\right|~,~ \\
 \label{eq:FT2}
 \Delta_{FT}(\mu_U) \approx&  \left|R \cdot 4 \sqrt{2} \frac{ \mu^2_U(M)}{v^2} \right|~,~
\end{align}
where $R = e^{((3 \log(M_{SUSY}/M) Y_t^2)/(16 \pi^2))}$, and  $\mu_U(M)$ is the running value of $\mu_U$ at the conformal scale. For simplicity we assumed that $Y_t$ does not change significantly between the SUSY breaking and conformal scales,
but our conclusion is independent of this approximation. As usual, one finds that the FT measure increases quickly  with
increasing values for the SUSY breaking scale and/or the scale of the dimensionful parameters in the superpotential. For $\mu_U$ in the TeV range, it seems not to be possible to find a FT measure below 100 unless the conformal scale is very low.

Assuming that all the parameters with dimension mass are correlated at the conformal scale,
which is defined as single scale supersymmetry, we have
\begin{eqnarray*}
\Lambda_D \sim \Lambda_F \sim \mu_D \sim \mu_U \sim m_s \sim \sqrt{B_\mu}~.~\,
\end{eqnarray*}
The underlying assumption is: there is one and only one fundamental parameter with dimension mass and
 the coefficients to set the different scales are calculable. However, a concrete construction of such a model
 is beyond the scope of this letter. As we will show, the single scale SUSY  condition can be relaxed further
 to the effective conformal sequestering single scale SUSY condition, and
for the following discussion only $\Lambda_F \sim \mu_U$ is necessary since their corresponding EWFT measures
are relatively large while all the rest are small and negligible.

\begin{figure}[hbt]
\includegraphics[width=0.9\linewidth]{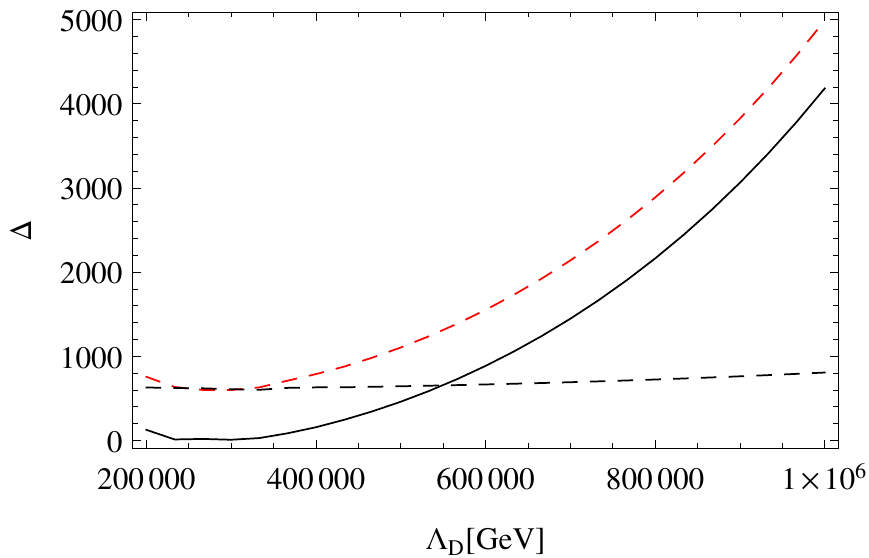} \\[5mm]
\includegraphics[width=0.9\linewidth]{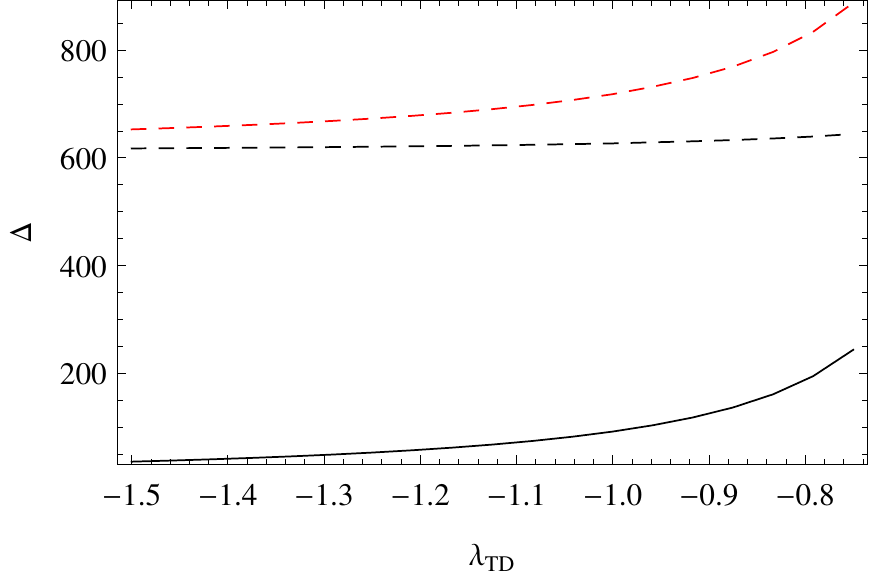} \\[5mm]
\includegraphics[width=0.9\linewidth]{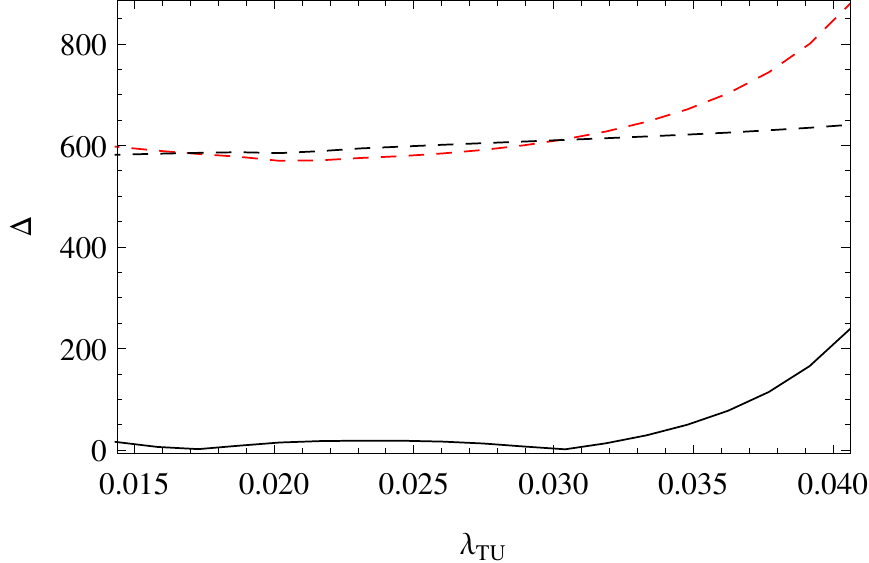}
\caption{The calculated FT measures $\Delta_{FT}(\mu_U)$ (dashed red), $\Delta_{FT}(\Lambda_F)$ (dashed black), $\Delta^C_{FT}$  (full black line) as function of $\Lambda_D$ (first row), $\lambda^T_d$ (second row) and $\lambda^T_u$ (third row). The other parameters were set to $\Lambda_F = 2.7\times 10^5$~GeV, $\Lambda_D = 2.0\times 10^5$~GeV, $M=10^{12}$~GeV, $\tan\beta=20$, $y_i=(0.6,-0.15,-0.75)$, $\lambda_{s,d} = (-0.78, -0.01)$, $\lambda^T_{s,d} = (-1,0.037)$, $v_s = -5$~GeV, $v_T=-0.25$~GeV.}
\label{fig:FT}
\end{figure}

To study the effect of this correlation, we first determine  $\mu_U(M)$ from the tadpole equations
in the limit $v_T \to 0$ and $\lambda_u\to0$. We obtain
\begin{align}
\label{eq:FT3}
& \mu_U(M) = \frac{1}{96 \pi^2 (\lambda^{T,2}_d-\lambda^{T,2}_u \tan^2\beta)} \Big(
6 g_2^2 \lambda^T_u \tilde \Lambda_D \tan^2\beta \nonumber \\
& + \sqrt{3 \lambda^{T,2}_d \tan^2\beta \left(12 g_2^4 \tilde \Lambda_D^2  +\lambda^{T,2}_u \Lambda_F^2 R\right) -3 \lambda^{T,4}_d \Lambda_F^2 R} \Big), \nonumber
\end{align}
where $\tilde \Lambda_D \equiv y_2 \Lambda_D$. If we combine Eqs.~(\ref{eq:FT1}) and (\ref{eq:FT2}),
we get the correlated FT measure
\begin{align}
& \Delta^C_{FT} = \frac{\sqrt{2} \lambda^{T,2}_u \Lambda_F^2  \tan^2\beta \, R}{384 \pi^4 v^2 \left(\lambda^{T,2}_d-\lambda^{T,2}_u \tan^2\beta\right)} + \frac{\tilde \Lambda_D}{\Lambda_F} F_1 +  \frac{\tilde \Lambda^2_D}{\Lambda^2_F} F_2 \,, \nonumber
\end{align}
where $F_1$ and $F_2$ are functions of $g_2$, $\lambda^T_i$ and $\tan\beta$ which we skip for brevity.
The last two terms can be suppressed in the limit $ \tilde \Lambda_D \ll \Lambda_F$. This is also the preferred limit,
because large $\tilde \Lambda_D$ would cause large wino masses which reduce the tree-level Higgs mass.
The first term becomes very small for $\lambda^T_u \to 0$. We have checked numerically that these estimates
reproduce the correct behaviour to a large extent even if we include the correlation to all other
dimensionful parameters. For this purpose, we implement the model in the \Mathematica package
\SARAH \cite{Staub:2008uz,Staub:2009bi,Staub:2010jh,Staub:2012pb,Staub:2013tta}
and generate Fortran code for \SPheno \cite{Porod:2003um,Porod:2011nf} to calculate the FT measures
using the full two-loop renormalization group equations (RGEs) based on Ref.~\cite{Goodsell:2012fm}. The calculated values
for $\Delta_{FT}(\Lambda_F)$,
$\Delta_{FT}(\mu_U)$ and $\Delta^C_{FT}$ as function of $\Lambda_D$, $\lambda^T_d$,
and $\lambda^T_u$ are shown in Fig.~\ref{fig:FT}.
We find that $\Delta^C_{FT}$ tends to be very small for small $\Lambda_F$ and $\lambda^T_u$ together with large $\lambda^T_d$, while $\Delta_{FT}(\Lambda_F)$ and
$\Delta_{FT}(\mu_U)$ are several orders larger.

Our proposal is completely different from focus point SUSY often considered
in the MSSM~\cite{Feng:1999mn, Feng:1999zg}: in the focus point SUSY $m_{H_u}^2$ is rather insensitive to the UV parameters because of specific hierarchies in the corresponding $\beta$-function. While this suppresses the FT with respect to  $m_{H_u}^2$ one has always to tune $\mu$ to be  small in order to obtain a low overall FT. In our proposal, there is no need that the FT with respect to the $\mu$-term is small nor the cancellations in the running of $m_{H_u}^2$ are needed because there is
a precise cancellation among these two sources.

We have checked whether this mechanism can be applied to the MSSM with the minimal GMSB. And indeed, we
have found there a good cancellation for large $\tan\beta$
if we relate the SUSY breaking scale $\Lambda$ and $\mu$ at the messenger scale.
However, this cancellation in the MSSM is not as good as the MRSSM considered here. The point is that the contributions
from the Majorana gaugino masses to the running of $m_{H_u}^2$ are absent in the MRSSM.
The $\Delta^{C}_{FT}$ in the MSSM is always bigger than the MRSSM, but still very good and well below 100.
We test it numerically by removing the gaugino contribution terms ``by hand'' from the $\beta$-functions of scalars,
and indeed
we can recover a similar cancellation as described here and  $\Delta^{C}_{FT}$ drops to very small values.
Thus, the supersoft character of Dirac gauginos together with an underlying correlation among
dimensionful parameters results in a very natural  model. The detailed study
for the MSSM will be given elsewhere.

%\section{Benchmark scenario}
%\label{sec:BP}

{\bf Benchmark Scenarios}--In the EWFT discussion  so far,
we have neglected all other current experimental constraints that must be fulfilled.
In particular, the  mass limits on the SUSY particles from direct and indirect searches as well as the measurement
of the SM-like Higgs mass exclude large parameter regions in SUSY models today. We can use the generated \SPheno version
to check all these constraints. It is especially worth to point out that the Higgs mass is also calculated
at the two-loop level including all model specific contributions in the gaugeless limit \cite{Goodsell:2014bna,Goodsell:2015ira}. Therefore, the theoretical uncertainty is of the same level as in the MSSM and can be estimated to be $O(3~\text{GeV})$.
We show the input and the most important output parameters
for two benchmark scenarios in Table~\ref{tab:BP}.

\begin{table}[h]
\begin{tabular}{|c|c|c|}
\hline
 & BP1 & BP2 \\
\hline
 \multicolumn{3}{|c|}{Input}                       \\
\hline
$\Lambda_F~[10^5~\GeV]$ &  2.7               &  2.0   \\
$\Lambda_D~[10^5~\GeV]$ &  2.0               &  2.2   \\
$M [10^7~\GeV]$         &  1.0               &  1.0   \\
$y_i$                   & -(-0.63,0.15,0.75) & -(0.45,0.16,1.1)    \\
$\lambda_{d,u}$         & -(0.78,0.01)      & (-1.45,0.09)    \\
$\lambda^T_{d,u}$       & (-1.0, 0.037)      & (-1.60,0.07)    \\
$\tan\beta$             &  20                &  20   \\
$v_{s,T}~[\GeV]$        & -(5.0,0.25)      & -(4.0,0.56)    \\
\hline
 \multicolumn{3}{|c|}{Output}                    \\
\hline
$\mu_U~[\GeV]$          & 1850      &     \\
\hline
 \multicolumn{3}{|c|}{Masses}                   \\
\hline
$m_h~[\GeV]$                  &  123.5     & 122.4    \\
$m_{\tilde g}~[\GeV]$         & 1620.5     &2316.9     \\
$m_{\tilde q}~[\GeV]$         &$\sim$ 3000 & $\sim$ 2300    \\
$m_{\tilde l_R}~[\GeV]$       &$\sim$ 500  & $\sim$400    \\
$m_{\tilde l_L}~[\GeV]$       &$\sim$1000  & $\sim$1000    \\
$m_{\tilde \chi_1^0}~[\GeV]$  &151.2       &  159.0   \\
\hline
 \multicolumn{3}{|c|}{$\Delta_{FT}$}            \\
\hline

Max($\Delta_{FT}(\lambda)$) & 0.7         & 1.8    \\
$\Delta_{FT}(\Lambda_F)$ &   342.5        & 180.0    \\
$\Delta_{FT}(\Lambda_D)$ &   0.2          & 0.1    \\
$\Delta_{FT}(\mu_U) $    &   342.8        & 186.7    \\
$\Delta_{FT}(\mu_D) $    &   4.2          &  9.2   \\
$\Delta_{FT}(B_\mu) $    &   4.3          &  9.1   \\
\hline
 \multicolumn{3}{|c|}{$\Delta^C_{FT}$}          \\
\hline
$\Delta^C_{FT}$          &    0.2          &  6.8  \\
\hline
\end{tabular}
\caption{The input parameters, important particle spectra, and EWFT measures
  for two benchmark scenarios. Max($\Delta_{FT}(\lambda)$) is the maximal EWFT measure
  for $\lambda_{d,u}$ and $\lambda^T_{d,u}$.}
\label{tab:BP}
\end{table}

One sees that the uncorrelated EWFT measures in our model are already smaller than
in  the usual MSSM with GMSB. 
The reason are the additional loop corrections
which weak the need for very heavy stops significantly. For example,
 we have very large $\lambda_d$ couplings for BP2. This is similar to
the MSSM extensions with vector-like (s)tops where the additional loop corrections
cause an significant improvement in the EWFT measure~\cite{Nickel:2015dna}. Also 
the two-loop corrections are enhanced due to the presence of scalar octets.
Moreover, the correlated EWFT becomes
 much smaller due to the precise cancellation between the contributions from $\Lambda_F$ and $\mu_U$.
For BP1 the resulting EWFT is even smaller for the dimensionless parameters.
Because of the slightly larger value of $\lambda^T_{u}$,  as expected,
the cancellation for BP2 is not working
as good as for BP1, although $\Delta^C_{FT}$ is still very small.

{\bf Conclusion}--We considered the GMSB with conformal sequestering,
and found  that the naive EWFT measures in the MRSSM are similar to the other SSMs
except the minor improvements due to supersoft property
and additional loop contributions to the Higgs boson mass.
With the effective single scale SUSY condition that all dimensionful
parameters with large EWFT measures are correlated, we showed explicitly
an excellent cancellation between the
  dominant terms that contribute to the EWFT. As we expected,
  the correlated EWFT measure is of unit order even
  for the TeV-scale supersymmetric particle masses.

%\section*{Acknowledgements}

{\bf Acknowledgements}--We thank Jessica Goodman for very useful discussions.
This research was supported in part by the Natural Science Foundation of China
under grant numbers 11135003, 11275246, and 11475238 (TL).

\bibliography{MRSSM}
\bibliographystyle{h-physrev5}

\end{document}